\documentclass[prb,twocolumn,amsmath,amssymb,showpacs]{revtex4}

\usepackage{graphicx}

\begin{document}

\title{Poynting's theorem for planes waves at an interface: a scattering matrix approach}

\author{V. Dom\'{\i}nguez-Rocha}
\affiliation{Departamento de F\'{\i}sica, Universidad Aut\'onoma
Metropolitana-Iztapalapa, A. P. 55-534, 09340 M\'exico D. F., Mexico}

\author{C. Zagoya}
\affiliation{Departamento de F\'{\i}sica, Universidad Aut\'onoma
Metropolitana-Iztapalapa, A. P. 55-534, 09340 M\'exico D. F., Mexico}

\author{M. Mart\'{\i}nez-Mares}
\affiliation{Departamento de F\'{\i}sica, Universidad Aut\'onoma
Metropolitana-Iztapalapa, A. P. 55-534, 09340 M\'exico D. F., Mexico}

\begin{abstract}
We apply the Poynting theorem to the scattering of monochromatic electromagnetic planes waves with normal incidence to the interface of two different media. We write this energy conservation theorem to introduce a natural definition of the scattering matrix $S$. For the dielectric-dielectric interface the balance equation lead us to the energy flux conservation which express one of the properties of $S$: it is a unitary matrix. For the dielectric-conductor interface the scattering matrix, that we denote by ${\tilde S}$, is no longer unitary due to the presence of losses at the conductor. However, the dissipative term appearing in the Poynting theorem can be interpreted as a single absorbing mode at the conductor such that a whole $S$, satisfying flux conservation and containing ${\tilde S}$ and this absorbing mode, can be defined. This is a simplest version of a model introduced in the current literature to describe losses in more complex systems.
\end{abstract}


\maketitle


\section{Introduction}

The scattering matrix $S$ is a useful tool to describe the multiple scattering that occur when planes waves enter to a system which in general is of complex nature, like atomic nucleus, chaotic and/or disordered systems, etc. \cite{Newton,Mello} Although it is well known for particle waves in quantum mechanics \cite{Newton,Mello,Merzbacher,Cohen} it can also be applied to any kind of plane waves. In the electromagnetic context the transfer matrix \cite{Jackson,Reitz} is known instead of $S$ \cite{transfer1} but they are equivalent and related. \cite{Mello} By definition, $S$ relates the outgoing plane waves amplitudes to the incoming ones to the system, from which the {\em reflection} $R$ and {\em transmission} $T$ coefficients are obtained; they are called {\em reflectance} and {\em transmitance} in electromagnetism. \cite{Reitz} In the absence of dissipation, as happens in quantum electronic and some electromagnetic systems, $S$ becomes a unitary matrix, in particular $R+T=1$. However, if the system contains a dissipative medium, $S$ is no longer unitary, in fact it is a sub-unitary matrix and $R+T<1$, where the lack of unity is called {\em absorbance} in the electromagentic subject. \cite{Reitz}

We are concerned with the electromagnetic case. There, a natural definition of $S$ through one of its properties arises from the Poynting theorem. This theorem is a energy balance equation which in the simplest form, i. e. for linear and non dispersive media, is given by \cite{Jackson}
\begin{equation}
\label{ec:cenergia}
{\bf \nabla}\cdot{\bf S}_P({\bf r},t) + 
\frac{\partial u({\bf r},t)}{\partial t}=
-\mbox{Re}{\bf J}({\bf r},t)\cdot\mbox{Re}{\bf E}({\bf r},t),
\end{equation}
where 
\begin{equation}
\label{ec:SP}
{\bf S}_P({\bf r},t) = \mbox{Re}{\bf E}({\bf r},t)\times \mbox{Re} 
{\bf H}({\bf r},t)
\end{equation}
is the Poynting vector which gives the energy flux per unit area per unit time, and $u({\bf r},t)$ is the electromagnetic energy density 
\begin{equation}
\label{ec:u}
u({\bf r},t) = \frac 12[ \mbox{Re}{\bf E}({\bf r},t)
\cdot\mbox{Re}{\bf D}({\bf r},t) + 
\mbox{Re}{\bf B}({\bf r},t)\cdot\mbox{Re}
{\bf H}({\bf r},t) ].
\end{equation}
Here, we have assumed that the electric ${\bf E}({\bf r},t)$, electric displacement ${\bf D}({\bf r},t)$, magnetic induction ${\bf B}({\bf r},t)$, and magnetic ${\bf H}({\bf r},t)$ fields, as well as the current density 
${\bf J}({\bf r},t)$, are complex vectors whose real part only has physical meaning. We have also written explicitely the dependence on the position 
${\bf r}$ and time $t$. The term in the right hand side of Eq.~(\ref{ec:cenergia}) is the negative of the work done by the fields per unit volume and represents the conversion of electromagnetic energy to thermal (or mechanical) energy. The version of Poynting's theorem for dispersive media will not be touched here.

Our purpose in this paper is to apply the Poynting theorem to the simplest  scattering system, an interface between two different media, to illustrate the relation of Poynting's theorem and one property of $S$. First, we will consider the absence of dissipation for the dielectric-dielectric interface for which the defined $S$ matrix is unitary. Second, the dissipative case is considered in the dielectric-conductor interface for which the scattering matrix, called ${\tilde S}$, is sub-unitary. The last system is the  simplest example to explain quantitatively a model, that we call the ``parasitic channels'' model, introduced in contemporary physics \cite{Lewenkopf1992,Brouwer1997} to describe complex systems with losses. \cite{Fyodorov2005} Also, $\tilde{S}$ could represents the scattering of an  ``absorbing patch'' used to describe surface absorption. \cite{MM-M}

The paper is organized as follows. In the next section we write the time averaged Poynting's theorem and calculate the corresponding Poynting's vector for planes waves and its flux through an open surface in a dielectric medium, as to be used in the sections that follow Sect. \ref{sec:Poynting}. The dielectric-dielectric interface is considered in Sect. \ref{sec:dielectric} while Sect. \ref{sec:conductor} is devoted to the dielectric-conductor interface. Finally, we conclude in Sect. \ref{sec:conclusions} 


\section{Poynting's theorem for linear and non-dispersive media}
\label{sec:Poynting}

In what follows we will consider monocromatic high frequency oscillating fields such that 
\begin{eqnarray}
\label{ec:E}
{\bf E}({\bf r},t) & = & {\bf E}({\bf r})\, e^{-i\omega t} \\ 
\label{ec:H}
{\bf H}({\bf r},t) & = & {\bf H}({\bf r})\, e^{-i\omega t}, \\
\label{ec:J}
{\bf J}({\bf r},t) & = & {\bf J}({\bf r})\, e^{-i\omega t},
\end{eqnarray}
where have written the precise dependence on the spatial and temporal variables to avoid confusion due to the abuse of the notation. Here, the temporal average is of importance. Using the definition 
\begin{equation}
\label{ec:promedio}
\langle f(t)\rangle = \frac{1}{\tau} \int_0^{\tau} f(t)\, dt.
\end{equation}
for the average of a time dependent function $f(t)$ over a period $\tau$, the  average of Eq. (\ref{ec:cenergia}) can be written as
\begin{equation}
\label{ec:energy-prom}
\nabla\cdot \langle {\bf S}_P({\bf r},t)\rangle = 
-\langle 
\mbox{Re}{\bf J({\bf r},t)}
\cdot\mbox{Re}
{\bf E}({\bf r},t)
\rangle,
\end{equation}
where we have used that, with help of Eqs. (\ref{ec:E}) and (\ref{ec:H}), 
\begin{equation}
\int_0^{\tau}\frac{\partial}{\partial t}u({\bf r},t)\, dt = 
u({\bf r}, \tau) - u({\bf r}, 0) = 0.
\end{equation}
Also, by substitution of Eqs. (\ref{ec:E}) and (\ref{ec:H}) into Eq. (\ref{ec:SP}), it is easy to see that
\begin{equation}
\label{ec:proms}
\langle {\bf S}_P({\bf r},t) \rangle = 
\mbox{Re}\, {\bf S}_P({\bf r}), 
\end{equation}
with (we write again the precise dependence) 
\begin{equation}
\label{ec:sprom2}
{\bf S}_P({\bf r}) = 
\frac 12 {\bf E}({\bf r})\times {\bf H}^*({\bf r}), 
\end{equation}
being the time averaged Poynting's vector. In equivalent way, using Eqs. (\ref{ec:E}) and (\ref{ec:J}), we get
\begin{equation}
\langle \mbox{Re}{\bf J({\bf r},t)}
\cdot\mbox{Re}{\bf E}({\bf r},t) \rangle = 
\frac 12 \mbox{Re}
\left[
{\bf J}({\bf r})\cdot{\bf E}^*({\bf r})
\right].
\end{equation}
Finally, the time averaged energy flux conservation law is given by
\begin{equation}
\label{ec:cenergia2}
\nabla\cdot {\bf S}_P({\bf r}) = -\frac 12
{\bf J({\bf r})}\cdot{\bf E}^*({\bf r}),
\end{equation}
where only the real part has physical meaning. 

If we integrate over a volume $V$ enclosed by a close surface $\Sigma$, it says that the net flux $\Phi$ of ${\bf S}_P({\bf r})$ through $\Sigma$ is the negative of the work done by the fields if there are dissipative components in $V$. I. e., 
\begin{equation}
\label{eq:balance}
\Phi = -W, 
\end{equation} 
where, from one side, 
\begin{equation}
\label{eq:W}
W = \frac 12 \int_V {\bf J({\bf r})}\cdot{\bf E}^*({\bf r})\, dV
\end{equation} 
and, for the other side, using the divergence theorem $\Phi$ can be written as a surface integral over $\Sigma$ as
\begin{equation}
\label{eq:flux}
\Phi = \oint_{\Sigma}{\bf S}_P({\bf r})\cdot\hat{{\bf n}}da .
\end{equation} 

\subsection{Flux of Poynting's vector for plane waves through an open surface of area $A$}

For a monocromatic plane wave with linear polarization in the $x$ axis,  propagating along the positive $z$-direction, the spatial component is 
\begin{equation}
\label{ec:Eplana}
{\bf E}(z) = E_0 \, e^{ikz} \, \hat{{\bf x}} ,
\quad\mbox{with}\quad k = \frac{n\omega}c 
\end{equation}
being the wave number, $n$ the index of refraction of the medium and $c$ the speed of light in vacuum; $\hat{{\bf x}}$ is a unit vector pointing in the positive $x$-axis, and the complex number $E_0$ is the amplitude of 
${\bf E}(z)$. The magnetic field ${\bf H}(z)$, calculated from ${\bf E}(z)$ using one of the Maxwell equations (Faraday's law of induction), is 
\begin{equation}
\label{ec:Hplana}
{\bf H}(z)=\frac{n}{\mu_0 c}E_0 \, e^{ikz}\, \hat{{\bf y}},
\end{equation}
where $\hat{{\bf y}}$ is a unit vector pointing in the positive $y$-direction and $\mu_0$ is the permeability of the free space (we have assumed a non-magnetic medium). 

\begin{figure}
\includegraphics[width=0.8\columnwidth]{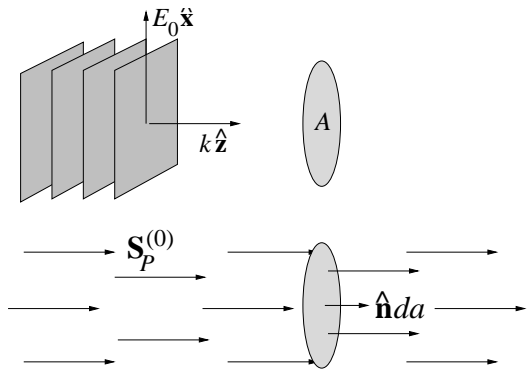}
\caption{A plane wave with linear polarization in the $x$ direction traveling in the $z$ direction cross the open surface of area $A$. The long arrows represent the corresponding time averaged Poynting's vector.}
\label{fig:flux}
\end{figure}

Sustituting Eqs.~(\ref{ec:Eplana}) and (\ref{ec:Hplana}) into Eq.  (\ref{ec:sprom2}) we get  
\begin{equation}
\label{ec:Splana}
{\bf S}^{(0)}_P(z) = \frac{n}{\mu_0 c}|{E_0}|^2\hat{{\bf z}},
\end{equation}
where $\hat{{\bf z}}$ is a unit vector pointing in the positive $z$-axis. 
This equation means that the time averaged energy flux is constant along the  propagation. If we consider an open surface of area $A$ as in Fig.~\ref{fig:flux}, the flux of ${\bf S}^{(0)}_P(z)$ through it is (see Eq.~(\ref{eq:flux})) 
\begin{equation}
\label{eq:fluxplane}
\Phi_0 = \int_A \frac{n}{\mu_0 c}|{E_0}|^2 da = \frac{n}{\mu_0 c}|{E_0}|^2 A,
\end{equation}
where we have taken the normal unit vector of $A$ as $\hat{\bf z}$. 


\section{Poynting's theorem at a dielectric-dielectric interface}
\label{sec:dielectric}

\begin{figure}
\includegraphics[width=0.8\columnwidth]{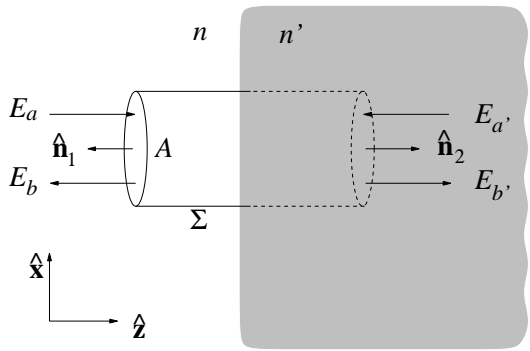}
\caption{Interface between two dielectrics with indices of refraction $n$ and $n'$. The amplitudes for the incoming ($E_a$ and $E_{a'}$) and outgoing ($E_b$ and $E_{b'}$) planes waves are shown with the corresponding Poynting's vectors represented by long arrows. Short arrows denote the unit vectors 
$\hat{\bf n}_1$ and $\hat{\bf n}_2$ of the covers of the cylindrical surface $\Sigma$. The cross section of the cylinder is $A$.}
\label{fig:energia}
\end{figure}

Let us consider an interface between two dielectrics, with refractive indices $n$ and $n'$, in the $xy$ plane as shown in Fig.~\ref{fig:energia}. For the 
shake of simplicity, we consider plane waves with linear polarization in the $x$ axis with normal incidence on both sides of the interface. Therefore, the spatial part of the electric fields, ${\bf E}(z)$ on the left and 
${\bf E}'(z)$ on the right, are of the form
\begin{eqnarray}
{\bf E}(z) & = & \left( E_a\,e^{ikz} + 
E_b\,e^{-ikz} \right) \hat{{\bf x}} \\
{\bf E}'(z) & = & \left( E_{a'}\,e^{ik'z} + 
E_{b'}\,e^{-ik'z} \right) \hat{{\bf x}} ,
\end{eqnarray}
where $E_a$ and $E_{a'}$ ($E_b$ and $E_{b'}$) are the complex amplitudes of the incoming (outgoing) planes waves.

For this case, the right hand side of Eqs.~(\ref{ec:cenergia2}) and (\ref{eq:balance}) is zero ($W=0$) because there is not dissipation. Hence, $\Phi=0$. For convenience, we take a cylinder of cross section $A$, including the covers, to be the closed surface $\Sigma$ (see Fig.~\ref{fig:energia}). Due to normal incidence only the flux through the covers of section $A$ contribute to $\Phi$. Then, Eq.~(\ref{eq:balance}) gives
\begin{equation}
\label{ec:cflujo}
\Phi_b + \Phi_a + \Phi_{b'} + \Phi_{a'} = 0, 
\end{equation}
where $\Phi_m$ ($m=a,\,a',\,b,\,b'$) is the flux through one of the covers with normal unit vector $\hat{{\bf n}}_j$ ($j=1$, 2) due to the $m$-th plane wave; it is given by
\begin{equation}
\label{ec:cflujo2}
\Phi_m = \int_A {\bf S}^{(m)}_P(z)\cdot\hat{{\bf n}}_j da.
\end{equation}
With the convention that $\hat{{\bf n}}_j$ points outwards of $\Sigma$,  $\Phi_b$ and $\Phi_{b'}$ are positive quantities while $\Phi_a$ and $\Phi_{a'}$ are implicitely negative. Using the result of  Eq.~(\ref{eq:fluxplane}) for a plane wave, Eq.~(\ref{ec:cflujo}) can be written as 
\begin{equation}
\label{ec:cflujo3}
n\left( |E_b|^2 - |E_a|^2\right) + 
n'\left( |E_{b'}|^2 - |E_{a'}|^2 \right) = 0,
\end{equation}
which can be arranged in a matricial form that we will use later, namely 
\begin{eqnarray}
\label{ec:cflujo4}
&&\left(\begin{array}{cc}
E^*_b & E^*_{b'}
\end{array}\right)
\left(\begin{array}{cc}
n & 0 \\ 0 & n'
\end{array}\right)
\left( \begin{array}{c}
E_b \\ E_{b'}
\end{array} \right) \nonumber \\ && =
\left(\begin{array}{cc}
E^*_a & E^*_{a'}
\end{array}\right)
\left(\begin{array}{cc}
n & 0 \\ 0 & n'
\end{array}\right)
\left( \begin{array}{c}
E_a \\ E_{a'}
\end{array} \right).
\end{eqnarray}

\subsection{The scattering matrix $S$ for the interface}
\label{sec:Sdiel}

By definition the Fresnel coefficients relate the outgoing to the incoming plane waves amplitudes as \cite{Jackson,Reitz,Marion}
\begin{equation}
\label{ec:Fresnel}
\left( \begin{array}{c}
E_b \\ E_{b'}
\end{array} \right) =
S_F \left( \begin{array}{c}
E_a \\ E_{a'}
\end{array} \right),
\end{equation}
where $S_F$ is the $2\times 2$ matrix 
\begin{equation}
\label{ec:Fresnel2}
S_F = 
\left(\begin{array}{cc}
-r_F & t'_F \\ t_F & -r'_F
\end{array}\right),
\end{equation}
The Fresnel coefficients of reflection $r_F$, $r'_F$ and transmission $t_F$,  $t'_F$ are \cite{Jackson} 
\begin{eqnarray}
\label{ec:cFresnel1}
r_F = \frac{n'-n}{n'+n}, & & r'_F = -r_F \\
\label{ec:cFresnel2}
t_F = \frac{2n}{n'+n}, & & 
t'_F = \frac{2n'}{n'+n}. 
\end{eqnarray}
Although for this particular case they are real numbers, in general they are complex, in which case $S_F$ is a complex matrix. In order to be more general, in what follows we will assume that $S_F$ is complex.  

By substitution of Eq.~(\ref{ec:Fresnel}), Eq.~(\ref{ec:cflujo4}) can be written as
\begin{eqnarray}
\label{ec:cflujo5}
\left(\begin{array}{cc}
E^*_a & E^*_{a'}
\end{array}\right)
S_F^{\dagger} \left(\begin{array}{cc}
n & 0 \\ 0 & n'
\end{array}\right)
S_F \left( \begin{array}{c}
E_a \\ E_{a'}
\end{array} \right) \nonumber \\ =
\left(\begin{array}{cc}
E^*_a & E^*_{a'}
\end{array}\right)
\left(\begin{array}{cc}
n & 0 \\ 0 & n'
\end{array}\right)
\left( \begin{array}{c}
E_a \\ E_{a'}
\end{array} \right), 
\end{eqnarray}
from which we see that 
\begin{equation}
S_F^{\dagger} \left(\begin{array}{cc}
n & 0 \\ 0 & n'
\end{array}\right) S_F = 
\left(\begin{array}{cc}
n & 0 \\ 0 & n'
\end{array}\right) 
\end{equation}
or, equivalently, 
\begin{equation}
\label{ec:Sflujo}
S^{\dagger}S = I, 
\end{equation}
where $I$ is the $2\times 2$ identity matrix and 
\begin{equation}
\label{ec:Sdef1}
S = \left(
\begin{array}{cc}
\sqrt{n} & 0 \\ 0 & \sqrt{n'}
\end{array}
\right)  S_F 
\left(
\begin{array}{cc}
\frac 1{\sqrt{n}} & 0 \\ 0 & \frac 1{\sqrt{n'}}
\end{array}
\right).
\end{equation}
$S$ is known in the literature as the {\em scattering matrix} which has the following general structure \cite{Mello}
\begin{equation}
\label{ec:Sdef2}
S = \left(
\begin{array}{cc}
r & t' \\ t & r'
\end{array}
\right).
\end{equation}
Here, $r$ and $t$ ($r'$ and $t'$) are the reflection and transmission  amplitudes for incidence on the left (right). From Eqs.~(\ref{ec:Fresnel2}), (\ref{ec:cFresnel1}), (\ref{ec:cFresnel2}) and (\ref{ec:Sdef1}) it is easy to see that 
\begin{eqnarray}
\label{eq:rrp}
r = \frac{n-n'}{n+n'}, & & r'=-r \\
\label{eq:ttp}
t = \frac{2\sqrt{nn'}}{n+n'} & & t'= t .
\end{eqnarray}
Two remarks are worthy of mention. \cite{Mello} The first one is that $t'=t$ because the system described by $S$ is invariant under inversion of time, but in a more general context that is not necessary the case. \cite{Dyson} The second is that $r'\neq r$ because the optical path is not the same for the reflected trajectory when incidence is from right or left.  
Of course, Eq.~(\ref{ec:Sflujo}) is satisfied in the case we are considering as can be easily checked using Eqs.~(\ref{eq:rrp}) and (\ref{eq:ttp}). In particular, 
\begin{equation}
\label{eq:unit1}
R+T=1,
\end{equation}
where $R=|r|^2$ and $T=|t|^2$ are the reflection and transmission coefficients. 

Eqs.~(\ref{ec:Fresnel}) and (\ref{ec:Sdef1}) implies that by definition $S$ relates the outgoing to incoming plane waves amplitudes but normalized with the index of refraction. I. e., 
\begin{equation}
\label{ec:Sdef3}
\left( \begin{array}{c}
\mathcal{E}_b \\ \mathcal{E}_{b'} 
\end{array} \right) =
S \left( \begin{array}{c}
\mathcal{E}_a \\ \mathcal{E}_{a'}
\end{array} \right),
\end{equation}
where $\mathcal{E}_i=\sqrt{n_i}E_i$ ($i=a,\,b,\,a',\,b'$), and the electric field on both sides is given by 
\begin{eqnarray}
{\bf E}(z) & = & 
\left(
\frac{\mathcal{E}_a}{\sqrt{n}}\, e^{ikz} + 
\frac{\mathcal{E}_b}{\sqrt{n}}\, e^{-ikz}
\right) \hat{{\bf x}} \\
{\bf E}'(z) & = & 
\left( 
\frac{\mathcal{E}_{a'}}{\sqrt{n'}}\, e^{ik'z} + 
\frac{\mathcal{E}_{b'}}{\sqrt{n'}}\, e^{-ik'z}
\right) \hat{{\bf x}}. \qquad
\end{eqnarray}

We recall that Eq.~(\ref{ec:Sdef3}) is a definition of $S$ that has arised in a natural way from the restriction imposed by flux conservation, Eq.~(\ref{ec:Sflujo}), and its  structure, given by Eq.~(\ref{ec:Sdef2}), reflects the symmetries present in the problem, one of them being the 
{\em time reversal invariance}. 


\section{The dielectric-conductor interface}
\label{sec:conductor}

\begin{figure}
\includegraphics[width=0.8\columnwidth]{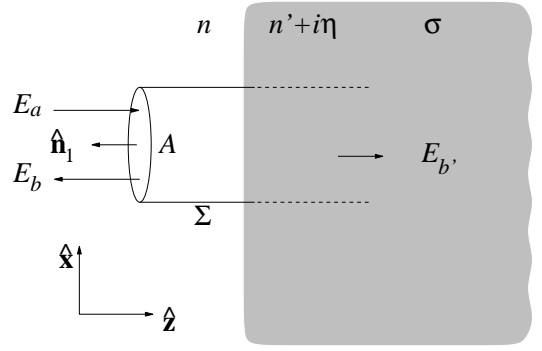}
\caption{The medium on the right is a conductor whose conductivity is $\sigma$. The refraction index, as well as the wave vector, becomes complex such that only evanescent waves are present with an amplitude $E_{b'}$ decaying exponentially as $z$ increases.}
\label{fig:energia2}
\end{figure}

When one of the two media is a conductor with electric conductivity $\sigma$, the one on the right in Fig.~\ref{fig:energia2} let say, the treatment is equivalent as in Sect. \ref{sec:dielectric} but with a complex index of refraction: $n'\rightarrow n'+i\eta$, where $n'$ and $\eta$ are the optical constants. \cite{Reitz} Also, the corresponding wave number becomes complex: we replace $k'\rightarrow k'+i\kappa$ with  
\begin{equation}
k' = n'\omega/c \quad \mbox{and} \quad \kappa = \eta\omega/c.
\end{equation}
Then, on the conductor side the electric field is an evanescent wave while on the dielectric side we have incoming and outgoing plane waves. They are 
\begin{eqnarray}
{\bf E}(z) & = & \left( E_a\,e^{ikz} + 
E_b\,e^{-ikz} \right) \hat{{\bf x}} \\
{\bf E}'(z) & = & E_{b'}\,e^{-\kappa z} \, e^{ik'z} \hat{{\bf x}}.
\end{eqnarray}
From Eq.~(\ref{ec:Fresnel}), or Eq.~(\ref{ec:Sdef3}), with $E_{a'}=0$ we see that 
\begin{eqnarray}
\label{eq:rF1x1} 
E_b & = & r\, E_a \\
\label{eq:tF1x1} 
E_{b'} & = & \sqrt{\frac n{n'+i\eta}}\,t\, E_a,
\end{eqnarray}
where $r$ and $t$ are obtained from Eqs.~(\ref{eq:rrp}) and (\ref{eq:ttp}) but replacing $n'$ by $n'+i\eta$, namely
\begin{eqnarray}
\label{eq:rrpcomplex}
r & = & \frac{n-n'-i\eta}{n+n'+i\eta},  \\
\label{eq:ttpcomplex}
t & = & \frac{2\sqrt{n(n'+i\eta)}}{n+n'+i\eta} .
\end{eqnarray}

In this case, the scattering matrix of the system is $1\times 1$ and we denote it by ${\tilde S}$. By definition (it is not necessary to normalize with respect to the index $n$) 
\begin{equation}
\label{eq:tildeS}
E_b = {\tilde S} \, E_a.
\end{equation}
Comparing with Eq.~(\ref{eq:rF1x1}) we see that  
\begin{equation}
\tilde{S} = r = \sqrt{R}\, e^{i\theta},
\end{equation}
where, from Eqs.~(\ref{eq:rrpcomplex}), 
\begin{equation}
\label{eq:Rtheta}
R = \frac{(n-n')^2 + \eta^2}
{(n+n')^2 + \eta^2}, 
\quad
\tan\theta = \frac{-2n\eta}{n^2-{n'}^2-\eta^2}.
\end{equation}

As for $S$, the Poynting theorem impose a  restriction to $\tilde{S}$. 
To apply Eq.~(\ref{eq:balance}) we consider that the closed surface $\Sigma$ in Fig.~\ref{fig:energia2} extends to infinity on the right of the interface, such that Eq.~(\ref{eq:W}) gives
\begin{equation}
W = \frac 12\int_0^{\infty}\sigma |E_{b'}|^2e^{-2\kappa z} Adz = 
\frac{\sigma}{4\kappa} |E_{b'}|^2A
\end{equation}
where we used that ${\bf J}(z)=\sigma {\bf E'}(z)$. Then, Eq.~(\ref{eq:balance}) can be written as 
\begin{equation}
\label{eq:Psub1}
\frac{n}{\mu_0 c}
\left( |E_b|^2 - |E_a|^2\right) = -\frac{\sigma}{2\kappa} |E_{b'}|^2
\end{equation}
where the area $A$ has been cancelled. Using Eqs.~(\ref{eq:tF1x1}) and (\ref{eq:tildeS}), after an arrangement, Eq.~(\ref{eq:Psub1}) gives 
\begin{equation}
\label{eq:Psub2}
|\tilde{S}|^2 + T_p = 1,
\end{equation}
meaning that $\tilde{S}$ is a $1\times 1$ subunitary matrix with $T_p$ the lack of unitarity ({\em strength of absorption} or absorbance), 
\begin{equation}
\label{eq:Tpdef}
T_p = \frac{\sigma\mu_0 c}{2\kappa\sqrt{{n'}^2+\eta^2}}|t|^2 = 
\frac{4nn'}{(n+n')^2+\eta^2}.
\end{equation}
The last equality is valid for metals in the upper infrared part of the spectrum and for metals at microwave and lower frequencies. \cite{Reitz-2}
Of course $T_p=1-|\tilde{S}|^2$ as can be easily verified using Eqs.~(\ref{eq:Rtheta}). 

Eq.~(\ref{eq:Psub2}) can be seen as resulting of the unitarity condition for an $S$-matrix that satisfy flux conservation (compare with  Eq.~(\ref{eq:unit1})). This $S$-matrix should has the structure (see Eq.~(\ref{ec:Sdef2}))
\begin{equation}
S = \left[
\begin{array}{cc}
\tilde{S} & \sqrt{T_p}\,e^{i\phi} \\ 
\sqrt{T_p}\, e^{i\phi} & \sqrt{1-T_p}\, e^{i(\phi-\theta)}
\end{array}
\right],
\end{equation}
where the phase $\phi$ can be taken as the phase of $\sqrt{n}t/\sqrt{n'+i\eta}$ (see Eqs.~(\ref{eq:tF1x1}) and (\ref{eq:Tpdef})); using Eq.~(\ref{eq:ttpcomplex}) 
\begin{equation}
\tan\phi = -\frac{\eta}{n'+n}.
\end{equation}

This form of $S$ says that the losses because of the conductor can be interpreted as due to a single mode of absorption whose ``coupling'' to the interface is $T_p$. This is the simplest version of what we call the ``parasitic channels'' model \cite{Lewenkopf1992,Brouwer1997} appeared in the literature of contemporary physics in recent years to describe power losses in more complex systems (see Ref. \onlinecite{Fyodorov2005} and references there in). That model consists in simulate losses with $N_p$ equivalent absorbing modes, each one having an imperfect coupling $T_p$ to the system. The total absorption is quantified by $\gamma=N_pT_p$ but $N_p$ and $T_p$ can not be determined separately. Our result not only explain this abstract model but quantify in an exact way the coupling $T_p$ of each absorbing mode as well the scattering matrix of a single absorbing patch in the surface absoprtion model. \cite{MM-M} Our treatment presented here can also be used to construct artificially a system with multiple absorbing modes. \cite{drmmc}


\section{Conclusiones}
\label{sec:conclusions}

We reduced the time averaged Poynting theorem to a property of the scattering matrix. For that we applied this balance equation to a simplest scattering system, consisting of normally incident planes waves at an interface between two media. The simplest version of this theorem was used such that dispersive media and the corresponding dissipation were ignored. Two kind of interfaces were considered. In the first one, a dielectric-dielectric interface, the energy flux conservation leads to a natural definition of the scattering matrix $S$ which is restricted to be a unitary matrix. We recalled that the structure of $S$ reflects the symmetries present on the problem in other contexts. In the second one, the dielectric-conductor interface, the definiton of $S$-matrix was used to describe the scattering taking into account the losses on the conductor side via the Poynting's theorem. This allowed us introduce the parasitic channels model used in contemporary physics to describe the scattering with losses in more complex systems. We should were able to quantify the coupling of the single parasitic mode of absorption in our simple system. A system with multiple absorbing modes can be constructed  and the results will be published elsewhere. Finally, the same treatement can be generalized for oblique incidence.

%
%
%

\end{document}